\documentclass[aps,amsmath,twocolumn,amssymb,floatfix,showpacs,superscriptaddress,nofootinbib,longbibliography]{revtex4-1}
\usepackage{mathtools}
\usepackage{braket}
\usepackage[dvipsnames]{xcolor}
\usepackage{float}
\usepackage{subfigure}
\usepackage[colorlinks=true,linktoc=page,linkcolor=BrickRed,citecolor=magenta,urlcolor=purple]{hyperref}
\newcommand{\vect}[1]{\boldsymbol{\mathrm{#1}}}
\mathchardef\mhyphen="2D 

\newcommand{\ie}{{\it i.e.,\,\,}}
\newcommand{\eg}{{\it e.g.,~}}

\newcommand\bea{\begin{eqnarray}}
\newcommand\eea{\end{eqnarray}}
\newcommand\beq{\begin{equation}}  
\newcommand\eeq{\end{equation}}

\newcommand{\non}{\nonumber}  
\usepackage[normalem]{ulem}
\definecolor{lime}{HTML}{A6CE39}
\usepackage{sidecap,tikz}
\DeclareRobustCommand{\orcidicon}{\hspace{-1.0mm}
	\begin{tikzpicture}
		\draw[lime, fill=lime] (0.0,0.0) 
		circle [radius=0.15] 
		node[white] {{\fontfamily{qag}\selectfont \tiny \,ID}};
		\draw[white, fill=white] (-0.0525,0.095) 
		circle [radius=0.007];
	\end{tikzpicture}
	\hspace{-3.0mm}
}
\foreach \x in {A, ..., Z}{\expandafter\xdef\csname orcid\x\endcsname{\noexpand\href{https://orcid.org/\csname orcidauthor\x\endcsname}
		{\noexpand\orcidicon}}
}

\AtBeginDocument{%
	\newwrite\bibnotes
	\def\bibnotesext{Notes.bib}
	\immediate\openout\bibnotes=\jobname\bibnotesext
	\immediate\write\bibnotes{@CONTROL{REVTEX41Control}}
	\immediate\write\bibnotes{@CONTROL{%
			apsrev41Control,author="08",editor="1",pages="1",title="1",year="1"}}
	\if@filesw
	\immediate\write\@auxout{\string\citation{apsrev41Control}}%
	\fi
}%
\begin{document}


\title{Second-order topological superconductor via noncollinear magnetic texture}  

\author{Pritam Chatterjee\orcidA{}}
\affiliation{Institute of Physics, Sachivalaya Marg, Bhubaneswar-751005, India}
\affiliation{Homi Bhabha National Institute, Training School Complex, Anushakti Nagar, Mumbai 400094, India}
\author{Arnob Kumar Ghosh\orcidB{}}
\affiliation{Institute of Physics, Sachivalaya Marg, Bhubaneswar-751005, India}
\affiliation{Homi Bhabha National Institute, Training School Complex, Anushakti Nagar, Mumbai 400094, India}
\affiliation{Department of Physics and Astronomy, Uppsala University, Box 516, 75120 Uppsala, Sweden}
\author{Ashis K. Nandy\orcidC{}}
\email{aknandy@niser.ac.in}
\affiliation{School of Physical Sciences, National Institute of Science Education and Research, An OCC of Homi Bhabha National Institute, Jatni 752050, India}
\author{Arijit Saha\orcidD{}}
\email{arijit@iopb.res.in}
\affiliation{Institute of Physics, Sachivalaya Marg, Bhubaneswar-751005, India}
\affiliation{Homi Bhabha National Institute, Training School Complex, Anushakti Nagar, Mumbai 400094, India}

\begin{abstract}
 We put forth a theoretical framework for engineering a two-dimensional (2D) second-order topological superconductor (SOTSC) by utilizing a heterostructure: incorporating noncollinear magnetic textures between an $s$-wave superconductor and a 2D quantum spin Hall insulator. It stabilizes the higher order topological superconducting phase, resulting in Majorana corner modes (MCMs) at four corners of a 2D domain. The calculated non-zero quadrupole moment characterizes the bulk topology. Subsequently, through a unitary transformation, an effective low-energy Hamiltonian reveals the effects of magnetic textures, resulting in an effective in-plane Zeeman field and spin-orbit coupling. This approach provides a qualitative depiction of the topological phase, substantiated by numerical validation within exact real-space model. Analytically calculated effective pairings in the bulk illuminate the microscopic behavior of the SOTSC. The comprehension of MCM emergence is supported by a low-energy edge theory, which is attributed to the interplay between effective pairings of $(p_x + p_y)$-type and $(p_x + i p_y)$-type. Our extensive study paves the way for practically attaining the SOTSC phase by integrating noncollinear magnetic textures.
\end{abstract}

\maketitle

\textcolor{blue}{\textit{Introduction.---}} The appearance of Majorana zero modes (MZMs) in topological superconductors (TSCs) has sparked significant interest in the quantum condensed matter community.~\cite{Kitaev2001,Ivanov2001,SDSarma2008,Kitaev2009,qi2011topological,Alicea_2012,Leijnse_2012,beenakker2013search}. 
In the quest to achieve MZMs in heterostructures, the placement of magnetic adatoms fabricated on a bulk $s$-wave superconductor presents a promising route, uniting theoretical and experimental efforts~\cite{Felix2013,AliYazdani2013,DanielLoss2013,PascalSimon2013,MFranz2013,Eugene2013,Felix2014,TeemuOjanen2014,MFranz2014,Rajiv2015,Sarma2015,Hoffman2016,Jens2016,Tewari2016,PascalSimon2017,Simon2017,Theiler2019,Cristian2019,Mashkoori2019,Menard2019,Pradhan2020,Teixeira2020,Alexander2020,Perrin2021,Nicholas2020,Debashish2023}.
The interplay between classical spin magnetism, represented by chains or adatoms, and superconductors (SCs) results in the emergence of Yu-Shiba-Rusinov (YSR) states (Shiba states), within the superconducting gap~\cite{Shiba1968,Felix2013,AliYazdani2013}. The overlap of Shiba states lead to Shiba bands, potentially governing a first-order TSC phase~\cite{Felix2013,AliYazdani2013,Kaladzhyan2016,Ojanen12015,Ojanen22016,Yong2022,Sebastian2022,Ghazaryan2022,Schmid2022,PChatterjee2023}, analogous to the one-dimensional Kitaev model~\cite{Kitaev2001,Kitaev2009}. Experimentally, the YSR states and/or the MZMs have been observed by growing magnetic impurities on an $s$-wave SC substrate~\cite{Eigler1997,Yazdani1999,Yazdani2015,Wiesendanger2021,Beck2021,Wang2021,Schneider2022,Richard2022,Wiesendanger2022,Yacoby2023,Soldini2023}. Nevertheless, the creation of YSR states goes beyond 1D systems; in a two-dimensional (2D) arrangement where noncollinear magnetic textures proximitized with an $s$-wave SC, unique effects like the emergence of 1D Majorana dispersive/flat edge modes emerges~\cite{Nagaosa2013,Ojanen12015,Ojanen22016,Balatsky2016,Jelena2016,Balatsky22016,Garnier2019,Rex2020,Dagotto2021,Tobias2022,Yong2022,Franz2022,chatterjee2023}, setting them notably different from the typical observation of MZMs.

\begin{figure}[]
\centering
\subfigure{\includegraphics[width=0.45\textwidth]{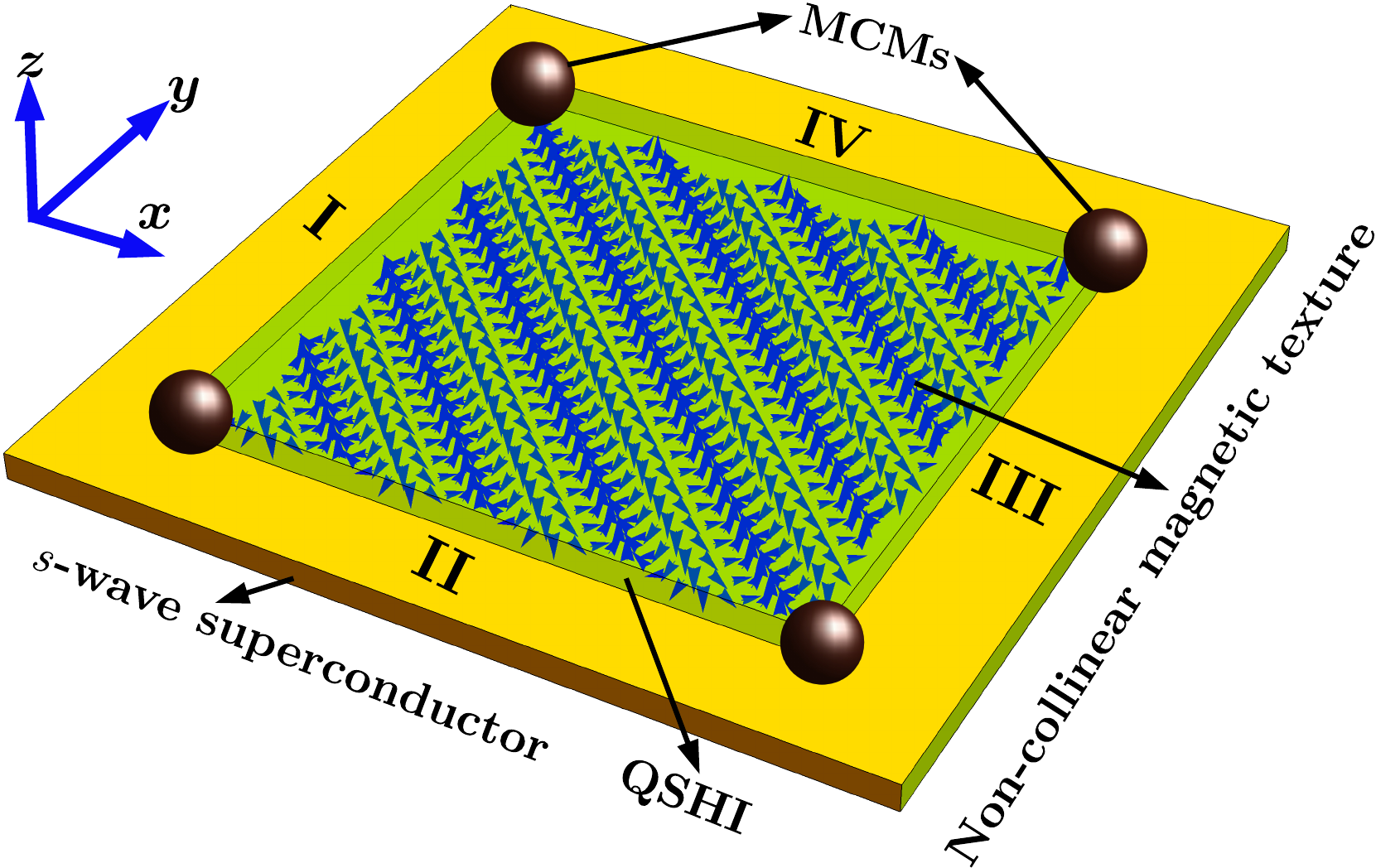}}
\caption{We present a schematic illustration of a heterostructure setup featuring a 2D noncollinear magnetic texture (blue arrows) and an $s$-wave superconductor (yellow), with a quantum spin Hall insulator (green) in between. In the topological phase, the system hosts localized Majorana corner modes (brown spheres) at its four corners. These features can be experimentally realized using standard scanning tunnelling microscopy (STM) technique. The four edges in the setup are labelled by I, II, III and IV.}
\label{Fig1}
\end{figure}
 
Conversely, the higher-order topological insulators~(HOTIs)~\cite{Hughes2017,Bernevig2017,Song2017,Langbehn2017,schindler2018,Franca2018,wang2018higher,Ezawakagome,Roy2019,Trifunovic2019,Khalaf2018,Ghosh2020,BiyeXie2021,trifunovic2021higher,Ghoshsystematic2022} and the higher-order topological superconductors~(HOTSCs)~\cite{Zhu2018,Franco2018,Zhong2018,Hughes2018,Fan2018,DasSarma2019,Klinovaja2019,Zhongbo2019,Enrico2019,Yan2019,Zhang2020,Daniel2020,Roy2020,Trauzettel2020,Ahn2020,Jiangbin2020,Ghosh2021-1,Frank2021,Xin2021,Ghosh2021,Bitan2021,GhoshHierarchy2021,Sarma2021,GhoshPRBL2022,Ghoshdynamical2022,PanPRB2022,GhoshPRB2023,Wong2023}, hosting $m$-dimensional boundary modes ($m$ \!=\! $d$\! --\! $n$, where $d$ is the dimension and $n$ is the topological order), have generated a profound research interest. 
In this emerging field, certain theoretical proposals offer elegant strategies to create second-order topological superconductors (SOTSCs) in 2D heterostructures hosting zero-dimensional (0D) Majorana corner modes (MCMs), involving 2D quantum spin Hall insulators (QSHIs) or two-dimensional electron gases with Rashba spin-orbit coupling (SOC) proximitized by an $s$-wave superconductor~\cite{Klinovaja2019,Zhang2020}. Notably, an in-plane Zeeman term stabilizes the MCMs. Another proposal for SOTSC centers around the ferromagnetic alignment of magnetic adatoms on an $s$-wave SC with Rashba SOC~\cite{Wong2023}. A SC version of the Bernevig-Hughes-Zhang (BHZ) model has also been proposed in the context of monolayer Fe(Se,Te) heterostructures~\cite{DasSarma2019}, where the magnetic layer exbibits bicollinear antiferromagnetic (AFM) order. Subsequently, Ref.~\cite{Soldini2023} introduces an alternative materials-centric strategy, presenting a sophisticated experimental plan to achieve the SOTSC phase, supported by a Rashba SOC-inclusive model Hamiltonian describing a magnet-superconductor hybrid (MSH) system. Indeed, the experiment has confirmed an AFM order of ${\rm Cr}$ layer on ${\rm Nb}(110)$. Currently, no theoretical proposal via a model Hamiltonian approach exists to realize the SOTSC phase in the presence of a noncollinear magnetic texture but excludes the Rashba SOC term. Thus, several intriguing questions arise to generate the SOTSC phase using a MSH setup: (a) How can a SOTSC phase hosting MCMs be achieved by initiating from a 2D-QSHI positioned between a texture of magnetic atoms and an $s$-wave SC? (b) What characterizes the pairing structures that are responsible for the emergence of the MCMs?

In this letter, our model setup comprises of a heterostructure geometry featuring a QSHI (mimicking a $\textrm{CdTe}$-$\textrm{HgTe}$-$\textrm{CdTe}$ type quantum-well~\cite{BHZ2006,Konig2007Science}) coupled with a noncollinear magnetic texture, positioned in close proximity to a bulk $s$-wave SC (see Fig.~\ref{Fig1}). Subsequently, we establish a unitary transformation to derive a momentum-space Hamiltonian, offering analytical insights into our analysis by calculating the effective low-energy edge theory. By employing a duality transformation, we uncover two varieties of SC pairings - $(p_x + p_y)$-type and $(p_x + i p_y)$-type - whose interplay leads to the emergence of the SOTSC phase. 

\begin{figure}[]
\centering
\subfigure{\includegraphics[width=0.49\textwidth]{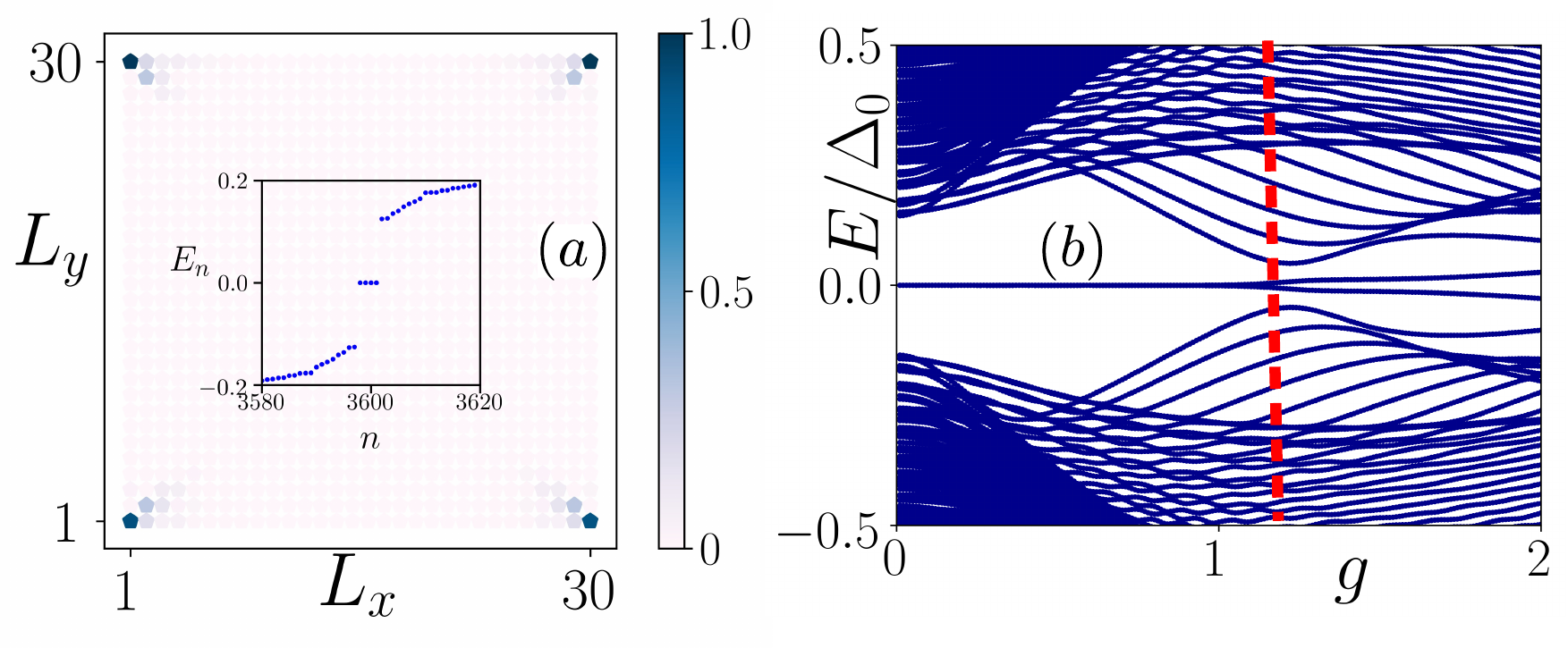}}
\caption{(a) The LDOS distribution associated with the energy $E$=$0$ is computed using $H$ [Eq.~(\ref{Eq1})] in the 2D domain \ie the $L_x \mhyphen L_y$ plane where MCMs are localized at four corners of the domain. In the inset, the eigenvalue spectrum as a function of the state index $n$ exhibits four zero-energy ($E$=0) modes within a gap. The simulated domain consists of $30 \times 30$ lattice sites while we choose the model parameters as, $t$=1.0, $J_{\rm ex}$=0.8, $g$=0.2, $\Delta_{0}$=0.4, $\lambda_x$=$\lambda_y$=$\lambda$=0.5, and $\epsilon_0$=1. In panel (b), the eigenvalue spectra corresponding to $H$ in Eq.~(\ref{Eq1}) are depicted as a function of the spiral wave vector $g$ under open boundary conditions. The topological phase transition point is highlighted by a 
vertical red dashed line.}
\label{Fig2}
\end{figure}
\vspace{0.1cm}
\textcolor{blue}{\textit{Realization of the SOTSC phase and characterizing its topological properties.---}} 
The real-space Hamiltonian for our configuration is given by:
\begin{widetext}
\vspace{-0.4cm}
\begin{align}
	\! H=  \! \sum_{i,j} \! c_{i,j}^\dagger \! \left[ \left\{ \epsilon_0 \Gamma_1 \! + \! \Delta_{0} \Gamma_2 \! + \! J_{\rm ex} \left(\Gamma_3 \cos \phi_{ij}  + \Gamma_4 \sin \phi_{ij} \right)  \right\} c_{i,j} \! - \!  \left\{  t \Gamma_1 +i \lambda_x \Gamma_5  \right\} c_{i+1,j} \! - \! \left\{  t \Gamma_1 +i \lambda_y \Gamma_6  \right\} c_{i,j+1} \right] + {\rm h.c.} \ ,
	\label{Eq1}
\end{align}
\vspace{-0.4cm}
\end{widetext}
where, the lattice site indices $i$ and $j$ runs along $x$- and $y$-direction, respectively and the $\vect{\Gamma}$ matrices ($8 \times 8$) are given as  $\Gamma_1$=$\tau_z \sigma_z s_0$, $\Gamma_2$=$\tau_x\sigma_0 s_0$, $\Gamma_3$=$\tau_0\sigma_{0}s_x$, $\Gamma_4$=$\tau_0\sigma_{0}s_y$, $\Gamma_5$=$\tau_z\sigma_x s_z$, and $\Gamma_6$=$\tau_z\sigma_y s_0$. The three Pauli matrices $\vect{\sigma}, \vect{s}$ and $\vect{\tau}$ act on orbital ($a, b$), spin ($\uparrow, \downarrow$), and particle-hole degrees of freedom, respectively. We work with the Bogoliubov-de Gennes~(BdG) basis as: $c_{i \in x,y}$=$\left\{\!c_{ia\uparrow},c_{ia\downarrow}, c_{ib\uparrow}, c_{ib\downarrow},-c^{\dagger}_{ia\downarrow}, c^{\dagger}_{ia\uparrow},  -c^{\dagger}_{ib\downarrow}, c^{\dagger}_{ib\uparrow\!}\right\}^T$, and $T$ denotes the transpose operation. Here, $\epsilon_0$, $\Delta_{0}$, $t$, and $\lambda_{x,y}$ represent staggered mass term, superconducting gap, hopping amplitude, and the strength of SOC, respectively. In this context, $J_{\rm ex}$ signifies the local exchange interaction strength between the magnetic impurity spin and the SC electrons, while $\phi$ denotes the angle between adjacent spins within the magnetic texture. We have chosen $\phi_{xy} = g_x x + g_y y$~\cite{chatterjee2023}, where the pitch of the noncollinear magnetic phase, particularly in the context of the spin-spiral state with a specific propagation direction, is dictated by $g_x$ and $g_y$. Note that, the Hamiltonian in Eq.~(\ref{Eq1}) reduces to the BHZ model of 2D QSHI~\cite{BHZ2006,Konig2007Science} when $J_{ex}=\Delta_0=0$. This model was proposed based on two specific types of materials, such as HgTe-CdTe quantum wells. These materials possess 
intrinsic SOC, represented by $\lambda_x$ and $\lambda_y$ in Eq.~(\ref{Eq1}). The BHZ model already exhibits first-order topology hosting gapless helical edge modes. This motivates us to achieve a SOTSC by introducing the terms $J_{ex}$ and $\Delta_0$, representing a noncollinear spin texture and an $s$-wave superconductor, respectively. The composite (three layer) system can then represent the schematic described in Fig.~\ref{Fig1} and the real space Hamiltonian introduced in Eq.~(\ref{Eq1}). To simplify matters, we assume $g_x$=$g_y$=$g$ and $\lambda_x$=$\lambda_y$=$\lambda$ in our numerical computations, without loss of generality. We qualitatively discuss our results in the case of rotational asymmetry, where $g_x \neq g_y$ and $\lambda_x \neq \lambda_y$ (see the supplemental material~(SM)~\cite{supp} for details). We further emphasize that higher-order topology predicted in our model does not depend on the square-shaped geometry, and one can still obtain 
the MCMs in a disc or triangular geometry (see SM~\cite{supp} for details).

Moving towards the numerical results associated with the Hamiltonian in Eq.~(\ref{Eq1}), we analyze the eigenvalue spectrum and the local density of states~(LDOS). We depict the LDOS associated with 
$E$=0 in Fig.~\ref{Fig2} (a) and in the inset, the eigenvalue spectrum $E_n$ as a function of state index $n$  is illustrated, obtained by utilizing open boundary conditions (OBC). 
The Majorana modes are located at $E=0$ with a negligible separation from the zero-energy $\sim \mathcal{O}(10^{-7})$ in a finite size system. The presence of zero-energy states becomes evident 
when examining the eigenvalue spectrum, and the localization of these states at the four corners of the 2D domain \ie the MCMs corroborates the second-order topological nature of the system~\cite{Ghosh2021-1}. To trace out the phase boundary, we display the eigenvalue spectra $E$ of $H$ as a function of the spiral pitch vector $g$ in Fig.~\ref{Fig2} (b) and here, we highlight a qualitative phase transition boundary, indicating a transition to a trivial SC state. Note that, in the limit $g$=$0$ (\ie a trivial collinear magnetic texture), $H$ in Eq.~(\ref{Eq1}) resembles the system with a QSHI, $s$-wave SC, and a constant Zeeman field as discussed in Refs.~\cite{Zhang2020,Ghosh2021-1}. On the other hand, we emphasize that in our system, we consider a spatial variation of the magnetic impurity spins through noncollinear 2D magnetic textures, while in earlier studies in this context, a rotating applied magnetic field has been considered without any spatial variation~\cite{Trauzettel2020,PanPRB2022}. Importantly, our model Hamiltonian setup works without any external magnetic field as far as the origin of MCMs is concerned. 

\begin{figure}[]
\centering
\subfigure{\includegraphics[width=0.48\textwidth]{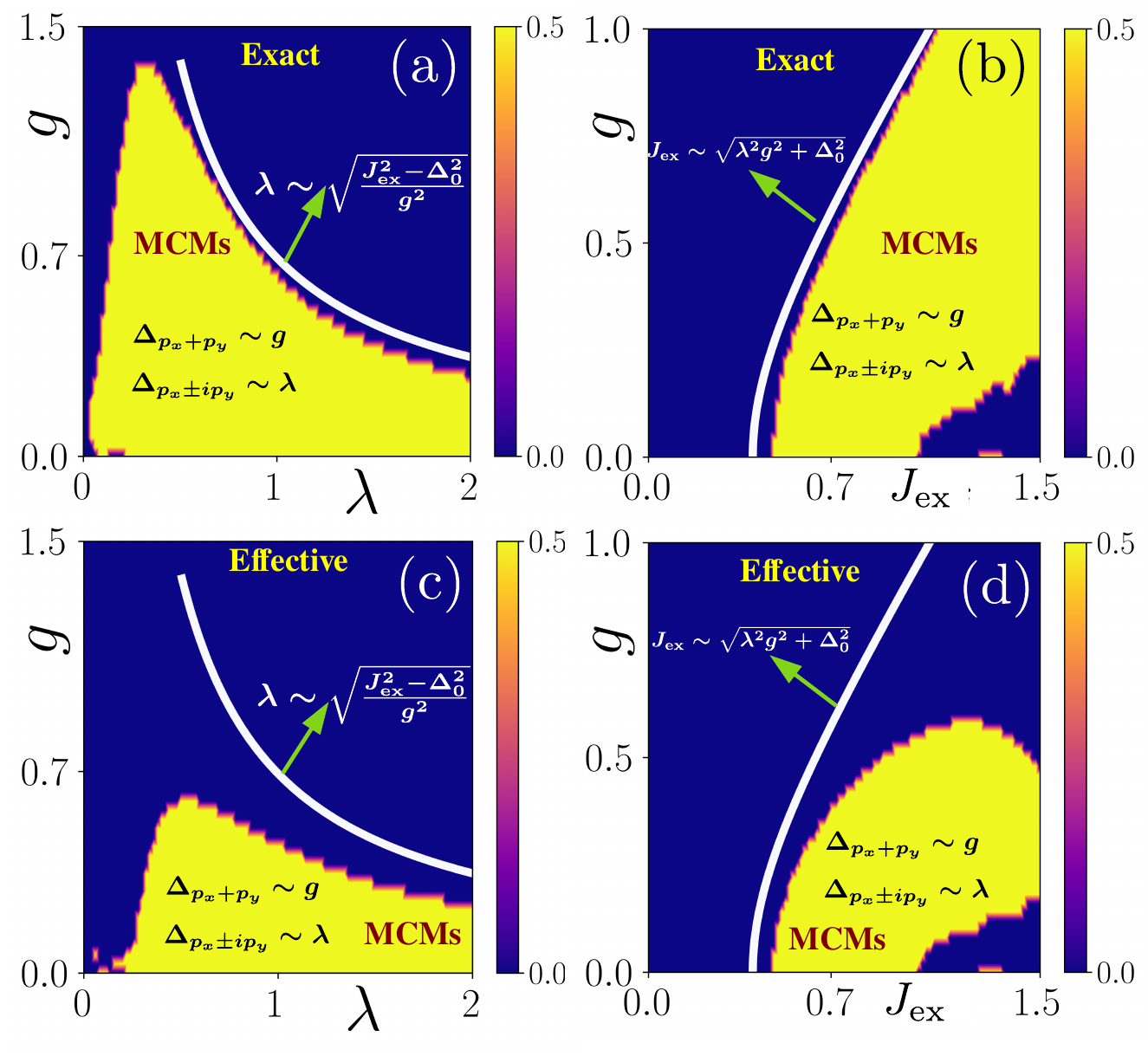}}
\caption{To establish the topological phases, the calculation of the quadrupole moment, $Q_{xy}$, is performed using both the exact real space Hamiltonian $H$ in Eq.(\ref{Eq1}) and the effective Hamiltonian in Eq.(\ref{Eq3}). For the first case, panels (a) and (b), illustrate the phases in the $\lambda \mhyphen g$ plane (with a fixed $J_{\rm ex}=0.8$) and  $J_{\rm ex} \mhyphen g$ plane (with a fixed $\lambda=1.0$), respectively. For the later case, these phase diagrams are depicted in panels (c) and (d). The calculated $Q_{xy}$ value is quantized to 1/2 within the topological region (yellow) and remains at zero in the trivial region (blue). In the lattice model, we consider a simulated domain of $30 \times 30$ lattice sites. The phase boundaries, indicated by the white line, is determined using the low-energy edge theory. The rest of the model parameters take the same value as mentioned in Fig.~\ref{Fig2}.}
\label{Fig3}
\end{figure}

The SOTSC phase can be topologically characterized by employing the bulk quadrupole moment ($Q_{xy}$) calculation~\cite{Hughes2017,Bernevig2017,Taylor2019,Kang2019}. Within the topological regime, the value of $Q_{xy}$ is quantized to 1/2 and hence, one expects the presence of highly localized corner states. In contrast, the value of $Q_{xy}$ becomes zero in the trivial phase. The $Q_{xy}$ is defined through the formula~\cite{Taylor2019,Kang2019,LiPRL2020}
\begin{equation}
	Q_{xy}= \frac{1}{2 \pi} {\rm Im} \left[ \ln \left\{\det \left(\mathcal{U}^\dagger \mathcal{W} \mathcal{U} \right) \sqrt{\det(\mathcal{W}^\dagger)} \right\}\right] \ ,
\end{equation}
where, $\mathcal{U}$ is an $N \!\times \! N_{\textrm {occ}}$ dimensional matrix encompassing the number of occupied eigenstates $N_{\rm occ}$ in $H$, see Eq.~(\ref{Eq1}), arranged according to their energy. The operator $\mathcal{W}$=$\exp \left[ i 2 \pi \hat{q}_{xy}\right]$ corresponds to the microscopic quadrupole operator $\hat{q}_{xy}$=$\hat{x} \hat{y} / L^2$, where $\hat{x}~(\hat{y}$) is position operator defined in the system with dimension $L$ along one direction.

By solving the lattice model Hamiltonian numerically, we calculate the quadrupole moment $Q_{xy}$ and illustrate its behavior in the $\lambda \mhyphen g$ and $J_{\rm ex} \mhyphen g$ plane in 
Figs.~\ref{Fig3}(a) and (b), respectively. Here, the yellow (blue) region designates a second order topological (trivial) regime with $Q_{xy}=0.5~(0)$. Fig.~\ref{Fig3}(a) indicates that one can obtain the SOTSC phase for $g\neq0$ in the presence of a nonzero SOC strength $\lambda$ in the QSHI. Moreover, an increase in the pitch $g$ reveals constraints on the permissible values of $\lambda$ necessary to exhibit the SOTSC phase. In Fig.~\ref{Fig3}(b), we observe that as the value of $g$ increases, there is a need to increase the local exchange interaction strength, $J_{\textrm ex}$ to achieve the topological phase. The phase boundaries in both Figs.~\ref{Fig3}(a) and (b) are bounded by a line (white lines), which one can compute analytically, and we provide that analysis in the latter part.

\textcolor{blue}{\textit{Effective model.---}} To add an analytical perspective to our investigation, we employ a low-energy continuum version of our Hamiltonian $H$ in Eq.~(\ref{Eq1}). Especially, we consider the following low-energy mixed-space Hamiltonian as,
\begin{equation}
H_{\rm L}=\xi_{\vect{k}} \Gamma_1 + \Delta_0 \Gamma_2 + 2 \lambda_x k_x \Gamma_5 +2 \lambda_y k_y \Gamma_6 
+J_{\rm ex}\vect{S}(\vect{r}).\vect{s}\ ,
\label{Eq2}
\end{equation}
where, $\xi_{\vect k}$=$\left[\epsilon_0-4t+t(k_x^2+k_y^2)\right]$ and the spin at position $\vect{r}$,  $\vect{S}(\vect{r})$=$\begin{pmatrix} \cos[\phi(\vect{r})],\!&\!\sin[\phi(\vect{r})],\!&\!0 \end{pmatrix}$. $\phi(\vect{r})$ quantifies the angle between neighboring spins. Here, $\vect{k}$=($k_x, k_y$) denotes the 2D momentum vector. We introduce a unitary transformation as $U$=$\tau_0 \sigma_0 e^{-\frac{i}{2}\phi(\vect{r})s_z}$, such that $H_{\text{eff}}=U^\dagger H_{\rm L} U$~\cite{chatterjee2023,Loss2022}. The effective Hamiltonian $H_{\rm eff}$ reads~(see the SM~\cite{supp} for the detailed derivation)
\begin{align}
H_{\text{eff}}=&\xi^{\text{eff}}_{\vect{k}} \ \Gamma_1 + \Delta_0 \Gamma_2 + J_{\rm ex} \Gamma_3 + 2 \lambda_x k_x \Gamma_5 +2 \lambda_y k_y \Gamma_6  \non \\
&-\frac{t}{2}(\vect{g} \cdot \vect{k}) \Gamma_7 -\lambda_xg_x \Gamma_8 - \lambda_yg_y \Gamma_9 \ ,
\label{Eq3}
\end{align}
where, $\xi^{\text{eff}}_k$=$\epsilon_0-4t + t(k_x^2\!+\! k_y^2)+(t/4)(g_x^2\!+\! g_y^2)$ and  $\Gamma_7$=$\tau_z\sigma_zs_z$, $\Gamma_8$=$\tau_z\sigma_xs_0$, and  $\Gamma_9$=$\tau_z\sigma_ys_z$. 
As a result of this transformation, the 2D noncollinear magnetic texture gives rise to an effective in-plane Zeeman field (proportional to $J_{\rm ex}$) and a corresponding effective spin-orbit coupling proportional to $\vect{g}$~\cite{chatterjee2023,Loss2022}.
Moreover, if we set $\lambda_{x,y}$=$0$, $H_{\rm eff}$ in Eq.~(\ref{Eq3}) for nonzero $g$ (=$\left|{\vect{g}}\right|$) gives rise to a gapless TSC phase hosting Majorana flat edge modes~\cite{chatterjee2023}. In a different scenario, if both $g_{x}$ and $g_{y}$ are set to zero while maintaining non-zero $\lambda_x$ and $\lambda_y$ values, Eq.~(\ref{Eq3}) closely resembles the 
model proposed in Ref.~\cite{Zhang2020} assuming the magnetization lies in-plane.

Utilizing this effective Hamiltonian, we demonstrate the topological phase diagram in the model parameter space as presented in Fig.~\ref{Fig3}(c) and (d), and we compare the same with the results obtained from the exact lattice model in Eq.~(\ref{Eq1}). In our approach, we employ the lattice regularized version of $H_{\textrm {eff}}$ by substituting $k_{x,y} \rightarrow \sin k_{x,y}$ and $1-(k_{x,y}^2/2) \rightarrow \cos k_{x,y}$, followed by the computation of $Q_{xy}$. Corresponding results in the $\lambda \mhyphen g$ and $J_{\textrm ex} \mhyphen g$ planes are shown in Figs.~\ref{Fig3}(c) and (d), respectively. While the effective theory successfully identifies the topological regions (depicted in yellow), it is important to note that the phase boundaries separating the topological and trivial phases deviate notably from the numerically obtained boundary lines using the exact lattice Hamiltonian, see the white lines in Figs.~\ref{Fig3}(c) and (d). Crucially, the effective theory provides a qualitative representation of the lattice model, demonstrating notably strong agreement for lower values of $g$. This discrepancy can be attributed to the fact that the effective Hamiltonian encapsulates only one component of the magnetic impurity spins ($J_{\rm ex} \Gamma_3$), while the other component gets suppressed during the transformation. While we initially consider a 2D noncollinear spin texture, the resulting effective magnetic field in this approach is confined to one direction. This infers that the complex magnetic spin texture might not be fully incorporated in the effective Hamiltonian, which could explain such deviation from the exact numerical result. Nevertheless, this $H_{\rm eff}$ in Eq.~(\ref{Eq3}) further sets the ground for us to investigate the emergence of the MCMs in our system, which we discuss in the subsequent text. Nevertheless, development of a formalism for the direct computation of $Q_{xy}$ from the low-energy effective Hamiltonian [Eq.~(\ref{Eq3})] can be an intriguing avenue 
for future investigation, and will be presented elsewhere.

\textcolor{blue}{\textit{Low-energy edge theory.---}} 
Here, we utilize the low-energy $H_{\rm eff}$ [Eq.~(\ref{Eq3})] to formulate an effective edge Hamiltonian for our 2D heterostructure geometry presented in Fig.~\ref{Fig1}. We employ the Fu-Kane's criteria such that $\epsilon^{\prime}$=$\big[ \epsilon_0\!-\! 4t \!+\! \frac{t}{4}(g_x^2+g_x^2) \big]\!\!<\!\!0$~\cite{Kane2007}. In the context of edge-II as a representative example within the 2D geometry in Fig.~\ref{Fig1}, we conduct the edge theory calculation.  Here, we apply periodic boundary condition along the $x$-direction and OBC along the $y$-direction.
Hence, we substitute $k_y$ with $-i\partial_y$ and treat $k_x$, $\Delta_0$, $J_{\rm ex}$, and $g_{x,y}$ as small parameters, ensuring that $tg_{x,y}$ remains finite. Neglecting the $k_x^2$ term, we can partition $H_{\rm eff}$ into $H_{\rm eff}=H_0(-i\partial_y)+H_p(k_x,\vect{g},J_{\rm ex},\Delta_{0})$, where their expressions are:
\begin{align}
    H_0=&(\epsilon^{\prime}-t \partial_y^2)\Gamma_1-2i\lambda_y\partial_y\Gamma_6+i\frac{tg_x}{2}\partial_y\Gamma_7 \ , \non \\
    H_p=&2\lambda_xk_x\Gamma_5-\frac{tg_x}{2}k_x\Gamma_7+\Delta_0 \Gamma_2+J_{\rm ex}\Gamma_3 -\lambda_xg_x\Gamma_8 \non \\
    &-\lambda_yg_y\Gamma_9 \ .
\end{align}
Through the exact solution of $H_0$ and perturbation calculations using the eigenstates of $H_0$, the matrix elements of $H_p$ are obtained; for detailed steps, see Section S2 of the SM~\cite{supp}.
The edge Hamiltonian associated with edge-II in Fig.~\ref{Fig1} is given by,
\begin{equation}
	\vspace{-0.15cm}
H_{\rm II}=2\lambda_x k_x \sigma_xs_z- \lambda_x g_x \sigma_xs_0 +J_{\rm ex}\sigma_0s_x +\Delta_{0}\sigma_zs_0 \ . 
\label{edge2Ham}
\end{equation}
This Hamiltonian  corresponds to a 1D Dirac equation with mass terms proportional to $\lambda_x g_x$, $\Delta_0$, and $J_{\rm ex}$. 
Following a similar procedure, one can procure the Hamiltonian for the other edges also~(see Section S2 of SM~\cite{supp} for a comprehensive discussion). We present a unified form for the 
Hamiltonian of all edges as follows:
\begin{equation}
    H_{j}= 2 A_j k_j \sigma_xs_z + B_j  \sigma_xs_0 +C_j \sigma_0s_x +\Delta_{0}\sigma_zs_0 \ , \label{edgeHamcombined}
\end{equation}
where, $j$ indicates edges as depicted by I, II, III, and IV in Fig.~\ref{Fig1}. Here, $k_\textrm{I,III}$=$k_y$ and $k_\textrm{II,IV}$=$k_x$, and $A_j$,$B_j$ and $C_j$ are $\left\{- \lambda_y, \lambda_x,- \lambda_y, \lambda_x \right\}$, $\left\{ \lambda_y g_y, -\lambda_x g_x, \lambda_y g_y, - \lambda_x g_x \right\}$, and $\left\{0, J_{\rm ex},0, J_{\rm ex} \right\}$, respectively. 

By examining the edge Hamiltonians presented in Eq.~(\ref{edgeHamcombined}), it is evident that the sign of the mass terms for the two intersecting edge Hamiltonians changes when 
$\sqrt{g^2\lambda^2+\Delta_{0}^2} >\! J_{ \rm ex}$ (see Section S2(C) in the SM~\cite{supp} for a detailed explanation). As a consequence, the Jackiw-Rebbi theory~\cite{Jackiw1975,Ghosh2021-1} can be employed to obtain the zero-energy modes that emerge at the intersection of the two edges, resulting in the formation of localized MCMs. Notably, the critical value of $g_c$, indicating the phase boundary for the emergence of MCMs, can be analytically computed from the gap-closing relation as $g_c$=$\sqrt{J_{\rm ex}^2 - \Delta_0^2} / \lambda$. We highlight this $g_c$ by a red dashed line in 
Fig.~\ref{Fig2} (b). Hence, the gap-closing relation offers an alternative analytical interpretation for the phase boundary indicated by the white line in Figs.~\ref{Fig3}(a)-(d), based on the values of $Q_{xy}$. The phase boundaries obtained from the analytical expressions align well with the actual boundaries observed in the numerical lattice model results, see Figs.~\ref{Fig3}(a) and (b). However, it exhibits notable discrepancies when compared to the boundaries derived from the lattice regularized version of the effective model as shown in Figs.~\ref{Fig3}(c) and (d).

\textcolor{blue}{\textit{Analysis of the effective bulk pairing.---}} 
After examining the topological phase boundaries and the emergence of MCMs, we briefly outline the nature of bulk superconducting pairing that is effectively generated through the interplay of SOC, the spin texture, and the $s$-wave SC utilizing the derived effective Hamiltonian in Eq.~(\ref{Eq3}). 
Applying a duality transformation~\cite{Satoshi2009,Satoshi2010}, we derive a dual Hamiltonian $\tilde{H}_{\rm D}$ from $H_\textrm{eff}$ as follows:
\begin{align}
    \! \tilde{H}_{\rm D} (\vect{k}) \! \! = \! U_{\rm D}^{\dagger}H_{\text{eff}}U_{\rm D} \! = \! \!
    \begin{pmatrix} 
    \!  \tilde{\epsilon}_{\rm D} \! + \! J_{\rm ex}\sigma_0 s_x & \tilde{\Delta}_{\rm D} (\vect{k}) \\
    \! \!\! \tilde{\Delta}_{\rm D} (\vect{k}) & \! \! \!\!  -\tilde{\epsilon}_{\rm D} \!+ \!J_{\rm ex}\sigma_0 s_x\\
\end{pmatrix},
\label{Eq9}
\vspace{-0.15cm}
\end{align}
where, $\tilde{\epsilon}_{\rm D}$=$\Delta_0\sigma_{0}s_0$ and $\tilde{\Delta}_{\rm D}$=$-\xi_{\vect{k}}^{\rm eff}\sigma_z s_0\!\!-\!\!\tilde{f}(\vect{k})$; with $\tilde{f}(\vect{k})$=$2\lambda_x k_x\sigma_x s_z \!+\!2\lambda_y k_y\sigma_y s_0\!-\!\lambda_x g_x\sigma_x s_0\!-\!\lambda_y g_y\sigma_y s_z \!-\!\frac{t}{2}(\vect{g}.\vect{k})\sigma_z s_z$~(see the details in Section S3 of SM~\cite{supp}). 
Despite of originating from $H_\textrm{eff}$ in Eq.~(\ref{Eq3}),  $\tilde{H}_{\rm D} (\vect{k})$ in Eq.~(\ref{Eq9}) does not inherently ensure the same topological phase as $H_{\rm eff}$~\cite{Satoshi2009,Satoshi2010}. Nevertheless, to ensure that the dual Hamiltonian is also capable of hosting the SOTSC phase, we compute $Q_{xy}$ by employing the real space formulation of 
$\tilde{H}_{\rm D} (\vect{k})$. Interestingly, $\tilde{H}_{\rm D} (\vect{k})$ also captures the same topological phase as $H_{\rm eff}$. 

Continuing our analysis, we examine the nature of bulk pairing using $\tilde{\Delta}_{{\rm D}} (\vect{k})$. 
Notably, we observe that the intrinsic SOC terms $\lambda_{x,y}$ undergo a transformation into $(p_x+ip_y)$-type SC pairings, yielding a gap $\Delta_{p_x+ip_y}$ proportional to  $\lambda$.
In contrast, the SOC induced due to the presence of noncollinear magnetic textures, characterized by $g_{x,y}$, transforms into a $(p_x+p_y)$-type SC pairing with a gap $\Delta_{p_x+p_y}$ that scales with $g$. In the context of 2D first-order TSC, it is well-known that $(p_x+ip_y)$-type pairing induces dispersive Majorana edge modes~\cite{Satoshi2009,Nagaosa2013,Hu_2019}, while $(p_x+p_y)$-type pairing is responsible for the emergence of flat Majorana edge modes~\cite{Nagaosa2013,Wang2017,Zhang2019,chatterjee2023}. When the SOC strength is set to zero, $\lambda_{x,y}$=$0$, the TSC 
phase exclusively features the flat edge modes~\cite{chatterjee2023}, which seems to restrict the emergence of opposite mass terms in the intersecting edge Hamiltonians. Thus, the presence of SOC is crucial in creating the dispersive edge modes, which can be gaped out by incorporating the sign-changing mass terms across the edge intersection. In conclusion, the system enters a SOTSC phase when the $(p_x+i p_y)$-pairing term ($\Delta_{p_x+ip_y}$) prevails over the $(p_x+ p_y)$-pairing term ($\Delta_{p_x+p_y}$). Moreover, when $\Delta_{p_x+ip_y}$ pairing is effectively induced in the edges, the corner modes are really of Majorana nature \ie $\tilde{\gamma}=\tilde{\gamma}^{\dagger}\Big\lvert_{E=0}$ where $\tilde{\gamma}$ is the quasi-particle operator.

\textcolor{blue}{\textit{Conclusion and Discussion.---}} 
We have investigated how the arrangement of a 2D heterostructure, featuring a QSHI sandwiched between a noncollinear magnetic texture and an $s$-wave superconductor, results in the SOTSC phase hosting MCMs. The numerical outcomes, encompassing the eigenvalue spectrum and the spatial distribution of the LDOS in a lattice model, demonstrate the feasibility of creating MCMs on a 2D finite domain. These features could be detectable using local probing methods, such as conventional scanning tunnelling microscopy (STM) experiments. We calculate $Q_{xy}$ using the lattice Hamiltonian to illustrate the phase diagram in different parameter space. Additionally, we establish a connection with an effective continuum model through a unitary transformation for analytical insights. The edge theory derived from the effective Hamiltonian can explain the numerically obtained MCMs. We also analyse the effective bulk SC pairings generated in this setup using a duality transformation. In summery, the SOTSC phase arises from the interplay of two types of pairings - ${p_x+ip_y}$ and ${p_x+p_y}$ - due to the presence of non-zero $\lambda$ and $\vect{g}$ in the system.

Regarding practical implementation, conventional SCs like ${\rm Nb}(110)$ exhibits a substantial SC gap, $\Delta_0 \approx$ 1.51 meV~\cite{Wiesendanger2021}. Recently, Mn/Nb(110) MSH system has found to exhibit the coexistence of AFM and SC phases together~\cite{Wiesendanger2022}. Nevertheless, the noncollinear spin-spiral state can be stabilized in heterostructures even with SC substrate, owing to the effects of frustration in exchange interactions~\cite{AKN_PRL2016,chatterjee2023}, and such systems can be fabricated and investigated using Scanning Tunneling Microscopy (STM) technique~\cite{Eigler1990,HowonScience2018,Schneider2020,Schneider2022}. Therefore, the potential experimental scenario for our setup involves placing a monolayer of magnetic adatoms (Mn/Cr etc.) on top of an $s$-wave superconductor (Nb/Al etc.). Such setup has made significant recent development for the experimental realization of first-order topological superconductivity, hosting Majorana zero modes~\cite{Wiesendanger2021,Schneider2022,Richard2022,Wiesendanger2022}. Additionally, in our theoretical proposal, we need to introduce another 2D layer of a QSHI, mimicking a $\textrm{HgTe}$-$\textrm{CdTe}$ quantum-well type structure~\cite{BHZ2006,Konig2007Science} for the realization of SOTSC hosting MCMs. To capture the signature of MCMs 
via LDOS [$\sim (dI/dV)$], one needs to employ a STM tip coated with either Nb or Cr on top of the trilayer heterostructure~\cite{Wiesendanger2022}. Considering the SC gap $\Delta_0\approx 1.51~ {\rm{meV}}$ (\eg Nb) as a reference, the remaining model parameters can take up the values (for Fig.~\ref{Fig2}):~$t \!\sim\!$ 3.78 meV, the magnetic impurity strength $J_{\rm ex}\sim$ 3.02 meV, the SOC $\lambda \!\sim\!$ 1.89 meV. However, for this set of other model parameter values, one can find from Fig.~\ref{Fig3} that the SOC strength $\lambda$ can take up a value as small as $0.23$ meV to realize the SOTSC phase. In literature, several articles have proposed potential values for the experimental parameters such as the intrinsic SOC strength ($\lambda$) for the material 
$\rm{Hg}_{0.32}\rm{Cd}_{0.68}\rm{Te}/\rm{HgTe}$~\cite{BHZ2006,Konig2007Science}, and exchange coupling strength ($J_{ex}\sim {\rm {eV}}$) for Fe/Mn etc.~\cite{Perge2014,Eigler1997}. 
In our theoretical work, we assume all these model parameters in terms of the superconducting gap $\Delta_{0}$. However, in a real experiment, the kinetic energy of the actual material is 100$-$1000 
times larger than the superconducting gap. Therefore, if we conduct a similar analysis on the kinetic energy (hopping element $t$) scale, the model parameters may fall within the range of possible materials.

\vskip -0.1cm
\textit{Acknowledgments.---} P.C., A.K.G., and A.S. acknowledge SAMKHYA: High-Performance Computing Facility provided by Institute of Physics, Bhubaneswar, for numerical computations. P.C. acknowledges Sandip Bera for stimulating discussions.  A.S. and A.K.N. acknowledge the support from Department of Atomic Energy (DAE), Govt. of India.

\bibliography{bibfile}{}

\clearpage
\newpage
\begin{onecolumngrid}
\begin{center}

{
\fontsize{12}{12}
\selectfont
\textbf{Supplemental material for ``Second-order topological superconductor via noncollinear magnetic texture''\\[5mm]}
}

\normalsize Pritam Chatterjee\orcidA{}$^{1,2}$, Arnob Kumar Ghosh\orcidB{}$^{1,2,3}$, Ashis K. Nandy\orcidC{}$^{4}$, and Arijit Saha\orcidD{}$^{1,2}$\\
{\small $^1$\textit{Institute of Physics, Sachivalaya Marg, Bhubaneswar-751005, India}\\[0.5mm]}
{\small $^2$\textit{Homi Bhabha National Institute, Training School Complex, Anushakti Nagar, Mumbai 400094, India}\\[0.5mm]}
{\small $^3$\textit{Department of Physics and Astronomy, Uppsala University, Box 516, 75120 Uppsala, Sweden}\\[0.5mm]}
{\small $^4$\textit{School of Physical Sciences, National Institute of Science Education and Research, An OCC of Homi Bhabha National Institute, Jatni 752050, India}\\[0.5mm]}

\end{center}

\normalsize
\newcounter{defcounter}
\setcounter{defcounter}{0}
\setcounter{equation}{0}
\renewcommand{\theequation}{S\arabic{equation}}
\setcounter{figure}{0}
\renewcommand{\thefigure}{S\arabic{figure}}
\setcounter{page}{1}
\pagenumbering{roman}

\renewcommand{\thesection}{S\arabic{section}}

In the supplementary material (SM), we provide detailed explanations in different sections. In Sec.~\ref{Sec:S1}, we explore the derivation of the effective Hamiltonian, which provides us with analytical insights into our numerical results. Moving to Sec.~\ref{Sec:S2}, we establish a low-energy edge theory based on our effective continuum model. Sec.~\ref{Sec:S3} is dedicated to deriving effective superconducting pairings in the bulk. Sec.~\ref{Sec:S4} and Sec.~\ref{Sec:S5} are devoted to the discussion of the effects of the asymmetry of the spin texture and Rashba spin-orbit coupling (SOC), respectively, on our setup. Finally, we discuss the emergence of Majorana corner modes~(MCMs) in disc and triangular geometry in Sec.~\ref{Sec:S6}.

\section{Derivation of the effective Hamiltonian } \label{Sec:S1}
In this section, we outline the method used to derive an effective Hamiltonian, $H_\textrm{eff}$, the Eq.(4) in the main text, from the low-energy continuum form of our lattice model $H$ \ie the Eq.(1) in the main text. In particular, we make use of the low-energy Hamiltonian $H_\textrm{L}$ as stated in Eq.~(3) of the main text, and upon replacing $(k_x, k_y)$ with $(-i\nabla_x,-i\nabla_y)$, it assumes the form:

\begin{equation}
H_{\rm L}=\xi \Gamma_1 + \Delta_0 \Gamma_2 + 2 \lambda_x k_x \Gamma_5 +2 \lambda_y k_y \Gamma_6 
+J_{\rm ex}\vect{S}(\vect{r}).\vect{s}\ ,
\end{equation}
where, $\Gamma_1=\tau_z\sigma_z s_0$, $\Gamma_2=\tau_x\sigma_0 s_0$, $\Gamma_3=\tau_0\sigma_{0}s_x$, $\Gamma_4=\tau_0\sigma_{0}s_y$, $\Gamma_5=\tau_z\sigma_x s_z$, $\Gamma_6=\tau_z\sigma_y s_0$, $\Gamma_7=\tau_z\sigma_zs_z$, $\Gamma_8=\tau_z\sigma_xs_0$, and  $\Gamma_9=\tau_z\sigma_ys_z$. Here, $\xi=\left[\epsilon_0-4t-t(\bigtriangledown_x^2+\bigtriangledown_y^2)\right]$. To obtain the effective continuum Hamiltonian, we introduce a unitary transformation $U=\tau_0 \sigma_0 e^{-\frac{i}{2}\phi(\vect{r})\sigma_z}$, such that $\tilde{H}_{\text{eff}}=U^\dagger H U$~\cite{chatterjee2023,Loss2022}. Hence, $\tilde{H}_{\text{eff}}$ reads as
\begin{align}
\tilde{H}_{\text{eff}}=&- t\sum_{r_i=x,y} \left[\nabla_{r_i}^2-\frac{1}{4}\left(\frac{\partial\phi}{\partial r_i}\right)^2-\frac{1}{2}\left(i\frac{\partial\phi}{\partial r_i}\nabla_{r_i} +i\nabla_{r_i}\frac{\partial\phi}{\partial r_i}\right)s_z\right]\sigma_z\tau_z +(\epsilon_0-4t)\Gamma_1+\Delta_0\Gamma_2+J_{\rm ex}\Gamma_3 +2\lambda_x k_x\Gamma_5 \non \\
&+2\lambda_y k_y\Gamma_6-\lambda_x\left(\frac{\partial\phi}{\partial x}\right) \Gamma_8-\lambda_y \left(\frac{\partial\phi}{\partial y}\right) \Gamma_9\ .
\end{align}
The effective Hamiltonian $H_\textrm{eff}$ takes on a specific form due to the presence of the 2D noncollinear spin texture defined as $\phi(x,y)=\vect{g}.\vect{r}=g_x x+g_y y$. As a result, the form of $H_\textrm{eff}$ becomes:
\begin{align}
H_{\text{eff}}=&\xi^{\text{eff}}_{\vect{k}} \ \Gamma_1 + \Delta_0 \Gamma_2 + J_{\rm ex} \Gamma_3 + 2 \lambda_x k_x \Gamma_5 +2 \lambda_y k_y \Gamma_6  -\frac{t}{2}(\vect{g} \cdot \vect{k}) \Gamma_7 -\lambda_xg_x \Gamma_8 - \lambda_yg_y \Gamma_9 \ .
\label{eq3}
\end{align}
The basis in which we perform our analysis is denoted as, 
\begin{align}
\Psi_{\vect{k}} = \left\{ c_{\vect{k},a,\uparrow},c_{\vect{k},a,\downarrow},c_{\vect{k},b,\uparrow},c_{\vect{k},b,\downarrow},-c^{\dagger}_{-\vect{k},a,\downarrow}, c^{\dagger}_{-\vect{k},a,\uparrow}, -c^{\dagger}_{-\vect{k},b,\downarrow},c^{\dagger}_{-\vect{k},b,\uparrow} \right\}^T.
\end{align}

\section{Low-energy edge theory from the effective model} \label{Sec:S2}
Here, we elaborate on the process of developing the low-energy edge theory follwoing Refs.~\cite{Zhong2018,Zhang2020,Ghosh2021-1} based on our effective continuum model presented in Eq.~(\ref{eq3}). We have examined the edges labeled as I and II, as illustrated in Figure 1 of the main text.

\subsection{Hamiltonian for edge-I}
For edge-I, we apply periodic boundary condition~(PBC) along the $y$-direction and open boundary condition~(OBC) along the $x$-direction, resulting in the substitution of $k_x$ with $-i\partial_x$ in the analysis. We divide the original Hamiltonian $H_\textrm{eff}$ into two terms: $H_{\text{eff}}=H_0(-i\partial_x)+H_p(k_y,\vect{g},J_{\rm ex},\Delta_{0})$, considering $H_p$ as a perturbation to $H_0$. By neglecting $k_y^{2}$ term in $H_0$, we obtain:
\begin{align}
	H_0=&(\epsilon^{\prime}-t \partial_x^2)\Gamma_1-2i\lambda_x\partial_x\Gamma_5+i\frac{tg_x}{2}\partial_x\Gamma_7\ , \non \\
	H_p=&2\lambda_yk_y\Gamma_6-\frac{tg_y}{2}k_y\Gamma_7+\Delta_0 \Gamma_2+J_{\rm ex}\Gamma_3 -\lambda_xg_x\Gamma_8 -\lambda_yg_y\Gamma_9 \ ,
\end{align}
where $\epsilon^{\prime}= \big[ \epsilon_0-4t+\frac{t}{4}(g_x^2+g_x^2) \big]$. Assuming the impact of $H_p$ on $H_0$ to be small, we solve $H_0$ exactly and treat $H_p$ as a perturbation. The trial solution for $H_0$ is assumed to be $\Psi_\alpha (x)=e^{\gamma x}\chi_\alpha$, satisfying $H_0\Psi_\alpha=0$ for the zero-energy states. Here, $\gamma$ is a complex quantity mimicking complex 
wave-vector and $\chi_{\alpha}$ represents eight-component spinors, see below.
Therefore, the secular equation reads,
\begin{equation}
	\det \left[(\epsilon^{\prime}-t\gamma^2)\Gamma_1-2i\lambda_x \gamma \Gamma_5+i\frac{tg_x}{2}\gamma\Gamma_7 \right]=0\ . 
\end{equation}
Employing the boundary condition $\Psi_\alpha(0)=\Psi_\alpha(\infty)=0$, the $\gamma_\alpha$ and $\chi_\alpha$ reads as
\begin{align}
	\gamma_1=&-\frac{-ig_x^{\prime}t+2\lambda_x+\sqrt{-(g_x^{\prime}t+2i\lambda_x)^2+4t\epsilon^{\prime}}}{2t}=-\left(\tilde{\alpha}+\frac{\lambda_x}{t}\right)- i\left(\tilde{\beta}-\frac{g_x^{\prime}}{2}\right)\ , \non \\
	\gamma_2=&\frac{ig_x^{\prime}t-2\lambda_x+\sqrt{-(g_x^{\prime}t+2i\lambda_x)^2+4t\epsilon^{\prime}}}{2t}=\left(\tilde{\alpha}-\frac{\lambda_x}{t}\right)+ i\left(\tilde{\beta}+\frac{g_x^{\prime}}{2}\right)\ , \non \\
	\gamma_3=&-\frac{ig_x^{\prime}t+2\lambda_x+\sqrt{-(g_x^{\prime}t-2i\lambda_x)^2+4t\epsilon^{\prime}}}{2t}=-\left(\tilde{\alpha}+\frac{\lambda_x}{t}\right)+ i\left(\tilde{\beta}-\frac{g_x^{\prime}}{2}\right)\ , \non \\
	\gamma_4=&\frac{-ig_x^{\prime}t-2\lambda_x+\sqrt{-(g_x^{\prime}t-2i\lambda_x)^2+4t\epsilon^{\prime}}}{2t}=\left(\tilde{\alpha}-\frac{\lambda_x}{t}\right)- i\left(\tilde{\beta}+\frac{g_x^{\prime}}{2}\right)\ ,
\end{align}
where $g_x^\prime=\frac{g_x}{2}$, $\tilde{\alpha}\!\!=\!\!\sqrt{\alpha^2+\beta^2}\cos (\theta/2)$, and $\tilde{\beta}\!\!=\!\!\sqrt{\alpha^2+\beta^2}\sin (\theta/2)$; with $\alpha = \frac{\epsilon^{\prime}}{t}+\frac{\lambda_x^2}{t^2}-\left(\frac{g_x^{\prime}}{2}\right)^2$, $\beta = \frac{g_x\lambda_x}{t}$, and $\theta = 2\tan^{-1}\!\!\left(\frac{g_x\lambda_x}{\epsilon^{\prime}-\frac{g_x^{\prime 2}t}{4}+\frac{\lambda_x^2}{t}}\right)$. The corresponding normalized spinors can be written as
\begin{equation} 
\chi_{1}= 
\frac{1}{2}
\begin{pmatrix*}[r]
1\\
0 \\
-i\\
0\\
1\\
0\\
-i\\
0
\end{pmatrix*}\ , \quad
\chi_{2}= 
\frac{1}{2}
\begin{pmatrix*}[r]
0\\
1\\
0\\
-i\\
0\\
1\\
0\\
-i
\end{pmatrix*}\ , \quad
\chi_{3}= 
\frac{1}{2}
\begin{pmatrix*}[r]
0\\
1\\
0\\
i\\
0\\
1\\
0\\
i
\end{pmatrix*}\ , \quad   
\chi_{4}= 
\frac{1}{2}
\begin{pmatrix*}[r]
1\\
0\\
i\\
0\\
1\\
0\\
i\\
0
\end{pmatrix*} \ .
\end{equation}
As a result, the matrix elements of $H_p$ using the aforementioned basis states are:
\begin{equation}
	H_{\alpha \beta}^{I,\rm Edge}=\int_{0}^{\infty} dx \ \Psi_{\alpha}^\dagger(x) H_p \Psi_{\beta}(x) \ . \
\end{equation}
We hence obtain the Hamiltonian for edge-I as,
\begin{equation}\label{edge1}
H_{\rm I}=-2\lambda_yk_y\sigma_xs_z+\Delta_{0}\sigma_zs_0+2\lambda_yg_y^\prime\sigma_xs_0\ .
\end{equation} 

Using a similar approach, the Hamiltonian for edge-III can be derived as follows:
\begin{equation}
H_{\rm III}=-2\lambda_yk_y\sigma_xs_z+\Delta_{0}\sigma_zs_0+2\lambda_yg_y^\prime\sigma_xs_0\ .
\end{equation}

\subsection{Hamiltonian for edge-II}
To derive the Hamiltonian for edge-II, we utilize PBC (OBC) along the $x$- ($y$)-direction, while also substituting $k_y$ with $-i\partial_y$ in the calculation  (see also the main text).
We now rewrite $H_{\text{eff}}$ as $H_{\text{eff}}=H_0(-i\partial_y)+H_p(k_x,\vect{g},J_{ex},\Delta_{0})$ and neglect the $k_x^{2}$ term. Thus, we obtain
\begin{align}
    H_0=&(\epsilon^{\prime}-t \partial_y^2)\Gamma_1-2i\lambda_y\partial_y\Gamma_6+i\frac{tg_x}{2}\partial_y\Gamma_7 \ , \non \\
    H_p=&2\lambda_xk_x\Gamma_5-\frac{tg_x}{2}k_x\Gamma_7+\Delta_0 \Gamma_2+J_{\rm ex}\Gamma_3 -\lambda_xg_x\Gamma_8 -\lambda_yg_y\Gamma_9 \ .
\end{align}
Assuming the trial solution of the Hamiltonian $H_0$ as $\Psi_\alpha=e^{\gamma y}\chi_\alpha$ and focusing on the zero-energy solution, we consider $H_0\Psi_\alpha=0$. Thus, the secular equation reads
\begin{equation}
    \det \left[ (\epsilon^{\prime}-t\gamma^2)\Gamma_1-2i\lambda_y\gamma \Gamma_6+i\frac{tg_x}{2}\gamma \Gamma_7 \right]=0 \ .
\end{equation}
Employing the boundary condition $\Psi_\alpha(0)=\Psi_\alpha(\infty)=0$, we obtain the $\gamma_\alpha$ and $\chi_\alpha$ as
\begin{align}
	\gamma_1=&-\frac{-ig_y^{\prime}t+2\lambda_y+\sqrt{-(g_y^{\prime}t+2i\lambda_y)^2+4t\epsilon^{\prime}}}{2t}=-\left(\tilde{\alpha}+\frac{\lambda_y}{t}\right)- i\left(\tilde{\beta}-\frac{g_y^{\prime}}{2}\right), \non \\
	\gamma_2=&\frac{ig_y^{\prime}t-2\lambda_y+\sqrt{-(g_y^{\prime}t+2i\lambda_y)^2+4t\epsilon^{\prime}}}{2t}=\left(\tilde{\alpha}-\frac{\lambda_y}{t}\right)+ i\left(\tilde{\beta}+\frac{g_y^{\prime}}{2}\right), \non \\
	\gamma_3=&-\frac{ig_y^{\prime}t+2\lambda_y+\sqrt{-(g_y^{\prime}t-2i\lambda_y)^2+4t\epsilon^{\prime}}}{2t}=-\left(\tilde{\alpha}+\frac{\lambda_y}{t}\right)+ i\left(\tilde{\beta}-\frac{g_y^{\prime}}{2}\right), \non \\
	\gamma_4=&\frac{-ig_y^{\prime}t-2\lambda_y+\sqrt{-(g_y^{\prime}t-2i\lambda_y)^2+4t\epsilon^{\prime}}}{2t}=\left(\tilde{\alpha}-\frac{\lambda_y}{t}\right)- i\left(\tilde{\beta}+\frac{g_y^{\prime}}{2}\right) \ ,
\end{align}
where $g_y^{\prime}\!\!=\!\!\frac{g_y}{2},~\tilde{\alpha}\!\!=\!\!\sqrt{\alpha^2+\beta^2}\cos (\theta/2)$, and $\tilde{\beta}\!\!=\!\!\sqrt{\alpha^2+\beta^2}\sin (\theta/2)$; with $\alpha = \frac{\epsilon^{\prime}}{t}+\frac{\lambda_y^2}{t^2}-\left(\frac{g_y^{\prime}}{2}\right)^2$, $\beta = \frac{g_y\lambda_y}{t}$, and $\theta = 2\tan^{-1}\!\!\left(\frac{g_y\lambda_y}{\epsilon^{\prime}-\frac{g_y^{\prime 2}t}{4}+\frac{\lambda_y^2}{t}}\right)$. The corresponding normalized spinors reads
\begin{equation} 
\chi_{1}= 
\frac{1}{2}
\begin{pmatrix*}[r]
1\\
0 \\
1\\
0\\
1\\
0\\
1\\
0
\end{pmatrix*}\ , \quad
\chi_{2}= 
\frac{1}{2}
\begin{pmatrix*}[r]
1\\
0\\
1\\
0\\
-1\\
0\\
-1\\
0
\end{pmatrix*}\ , \quad
\chi_{3}= 
\frac{1}{2}
\begin{pmatrix*}[r]
0\\
1\\
0\\
1\\
0\\
1\\
0\\
1
\end{pmatrix*}\ , \quad   
\chi_{4}= 
\frac{1}{2}
\begin{pmatrix*}[r]
0\\
1\\
0\\
1\\
0\\
-1\\
0\\
-1
\end{pmatrix*} \ .
\end{equation}
Therefore, the matrix elements of $H_p$ employing the above basis states can be written as,
\begin{equation}
	H_{\alpha \beta}^{II,\rm Edge}=\int_{0}^{\infty} dy \ \Psi_{\alpha}^\dagger(y) H_p \Psi_{\beta}(y) \ .
\end{equation}
Therefore, we obtain the Hamiltonian for edge-II as,
\begin{equation}\label{edge2}
H_{\rm II}=2\lambda_xk_x\sigma_xs_z+\Delta_{0}\sigma_zs_0+J_{\rm ex}\sigma_0s_x-2\lambda_xg_x^\prime\sigma_xs_0\ .
\end{equation}

In a similar fashion, one obtains the Hamiltonian for edge-IV as,
\begin{equation}	
H_{\rm IV}=2\lambda_xk_x\sigma_xs_z+\Delta_{0}\sigma_zs_0+J_{\rm ex}\sigma_0s_x-2\lambda_xg_x^\prime\sigma_xs_0\ .
\end{equation}

So, the Hamiltonians for all four edges are as follows:
\begin{align}
    H_{\rm I}=&-2\lambda_yk_y\sigma_xs_z+\Delta_{0}\sigma_zs_0+2\lambda_yg_y^\prime\sigma_xs_0\ , \non \\
    H_{\rm II}=&2\lambda_xk_x\sigma_xs_z+\Delta_{0}\sigma_zs_0+J_{\rm ex}\sigma_0s_x-2\lambda_xg_x^\prime\sigma_xs_0\ , \non \\
    H_{\rm III}=&-2\lambda_yk_y\sigma_xs_z+\Delta_{0}\sigma_zs_0+2\lambda_yg_y^\prime\sigma_xs_0\ , \non \\
    H_{\rm IV}=&2\lambda_xk_x\sigma_xs_z+\Delta_{0}\sigma_zs_0+J_{\rm ex}\sigma_0s_x-2\lambda_xg_x^\prime\sigma_xs_0\ .
\end{align}

\subsection{Condition for sign change of the mass gaps between the two edges}
We can compute the eigenvalues of the edge Hamiltonians $H_{\rm I}$ in Eq.~(\ref{edge1}) and $H_{\rm II}$ in Eq.~(\ref{edge2}) at $k_y=0$ and $k_x=0$, respectively,  leading to the following results:
\begin{align}
    E_{\rm I}=& \left\{ -\sqrt{\Delta_{0}^2+g_y^2\lambda_y^2}, \ -\sqrt{\Delta_{0}^2+g_y^2\lambda_y^2}, \ \sqrt{\Delta_{0}^2+g_y^2\lambda_y^2}, \ \sqrt{\Delta_{0}^2+g_y^2\lambda_y^2} \right\} \ , \non \\
    E_{\rm II}=& \left\{ -J_{\rm ex}-\sqrt{\Delta_{0}^2+g_x^2\lambda_x^2}, \  -J_{\rm ex}+\sqrt{\Delta_{0}^2+g_x^2\lambda_x^2}, \  J_{\rm ex}-\sqrt{\Delta_{0}^2+g_x^2\lambda_x^2}, \ J_{\rm ex}+\sqrt{\Delta_{0}^2+g_x^2\lambda_x^2}   \right\} \ .
\end{align}
For a non-zero value of $\Delta_0$, $g_y$, and $\lambda_y$, edge-I Hamiltonian always exhibits a gap at $k_y=0$. However, for edge-II, one can close the gap for $J_{\rm ex} = \lvert \sqrt{\Delta_{0}^2+g_x^2\lambda_x^2} \rvert$. Thus, comparing the eigenvalues determined from $H_{\rm I}$ and  $H_{\rm II}$, we obtain the condition for sign change of the mass gap between the two adjacent or crossing edges as
\begin{eqnarray}
	\sqrt{g^2\lambda^2+\Delta_{0}^2}>J_{\rm ex}\ .
\end{eqnarray}	
Here, we consider $g_x=g_y=g$ and $\lambda_x=\lambda_y=\lambda$ for simplicity. 
Hence, as per the Jackiw-Rebbi theory~\cite{Jackiw1975}, we identify the emergence of localized Majorana zero modes at the corners of the system (MCMs), indicating the presence of a 
second-order topological superconducting phase.

\section{Derivation of the effective pairing} \label{Sec:S3}
In the following section, we present the derivation of the effective pairings in the bulk as discussed in the main text. We begin by reexpressing the effective Hamiltonian $H_{\rm eff}$ in Eq.~(\ref{eq3}) as,
\begin{align}
H_{\text{eff}}=\begin{pmatrix} 
    \xi_{\vect{k}}^{\text{eff}}\sigma_z s_0+\tilde{f}(\vect{k})+J_{\rm ex}\sigma_0 s_x & \Delta_{0}\sigma_{0}s_0 \\
    \Delta_{0}\sigma_{0}s_0 & -\xi_{\vect{k}}^{\text{eff}}\sigma_z s_0-\tilde{f}(\vect{k})+J_{\rm ex}\sigma_0 s_x\\
\end{pmatrix}, 
\label{eq31}
\end{align}
where, $\tilde{f}(\vect{k})=2\lambda_x k_x\sigma_x s_z+2\lambda_y k_y\sigma_y s_0-\lambda_x g_x\sigma_x s_0-\lambda_y g_y\sigma_y s_z-\frac{t}{2}(\vect{g} \cdot \vect{k})\sigma_z s_z$. Here, we have neglected the $k_{x,y}^2$ terms. We have derived a ``dual" Hamiltonian which is unitary equivalent to the $H_{\rm eff}$~\cite{Satoshi2009,Satoshi2010}. The duality transformation is facilitated by the unitary operator, $U_{\rm D}=\tilde{\tau} \sigma_0 s_0$, where $\tilde{\tau}=\frac{1}{\sqrt{2}}\begin{pmatrix*}[r]
		1 & -1 \\
		1 & 1\\
	\end{pmatrix*}$. The matrix $\tilde{\tau}$ obeys the following relations,
\begin{align}
	\tilde{\tau}^{\dagger}\tau_z\tilde{\tau}=&-\tau_x \non\\
	\tilde{\tau}^{\dagger}\tau_x\tilde{\tau}=&\tau_z .
	\label{eq7}
\end{align}
After the transformation, the dual Hamiltonian $\tilde{H}_{\rm D} (\vect{k})$ is expressed as:
\begin{equation}
	\tilde{H}_{\rm D} (\vect{k})=U_{\rm D}^{\dagger}H_{\text{eff}}U_{\rm D}=\begin{pmatrix} 
		\tilde{\epsilon}_{\rm D}+J_{\rm ex}\sigma_0 s_x & \tilde{\Delta}_{\rm D} \\
		\tilde{\Delta}_{\rm D} & -\tilde{\epsilon}_{\rm D}+J_{\rm ex}\sigma_0 s_x\\
	\end{pmatrix}\ , 
	\label{eq36}
\end{equation}
where, $\tilde{\epsilon}_{\rm D}=\Delta_0\sigma_{0}s_0$ and $\tilde{\Delta}_{\rm D}=-\xi_{\vect{k}}^{\text{eff}}\sigma_z s_0-\tilde{f}(\vect{k})$. Here, $\tilde{\epsilon}_{\rm D}$ represents the kinetic term 
and the off-diagonal element $\tilde{\Delta}_{\rm D}$ denotes the effective pairings that we have discussed in the main text. However, the diagonal kinetic term $\tilde{\epsilon}_{\rm D}$ is independent of momenta. Nevertheless, from the off-diagonal pairing term $\tilde{\Delta}_{\rm D}$, it is evident that the SOC terms undergo a transformation into $p_{x}+ip_{y}$-type superconducting pairing proportional to $\lambda_{x,y}$. On the other hand, the SOC generated via the noncollinear magnetic textures transforms into a $p_{x} + p_{y}$-type pairing that scales with $g_{x,y}$. 

\section{Impact of the rotational asymmetry } \label{Sec:S4}
In this section, we discuss the effect of asymmetry of the spin spiral ($g_x\neq g_y$) and intrinsic SOC ($\lambda_x\neq \lambda_y$) on the SOTSC phase. It is crucial to note that rotational symmetries 
are not essential to obtain the SOTSC phase hosting MCMs in our system. 
As a result, even if we consider asymmetry either in $g_x$ and $g_y$ or $\lambda_x$ and $\lambda_y$ the system still exhibits four MCMs at energy $E=0$. In Fig.~\ref{Fig1}(a), Fig.~\ref{Fig1}(b), and 
Fig.~\ref{Fig1}(c), we present the eigenvalue spectrum and the corresponding local density of states (LDOS) for three distinct cases: (a) $g_x\neq g_y, \lambda_x=\lambda_y$, (b) $g_x=g_y, \lambda_x\neq \lambda_y$ and (c) $g_x\neq g_y, \lambda_x\neq\lambda_y$.  Notably, in all of these cases, we consistently observe the presence of four Majorana corner modes at energy $E=0$.
\begin{figure}[H]
	\begin{center}
		\includegraphics[width=0.98\textwidth]{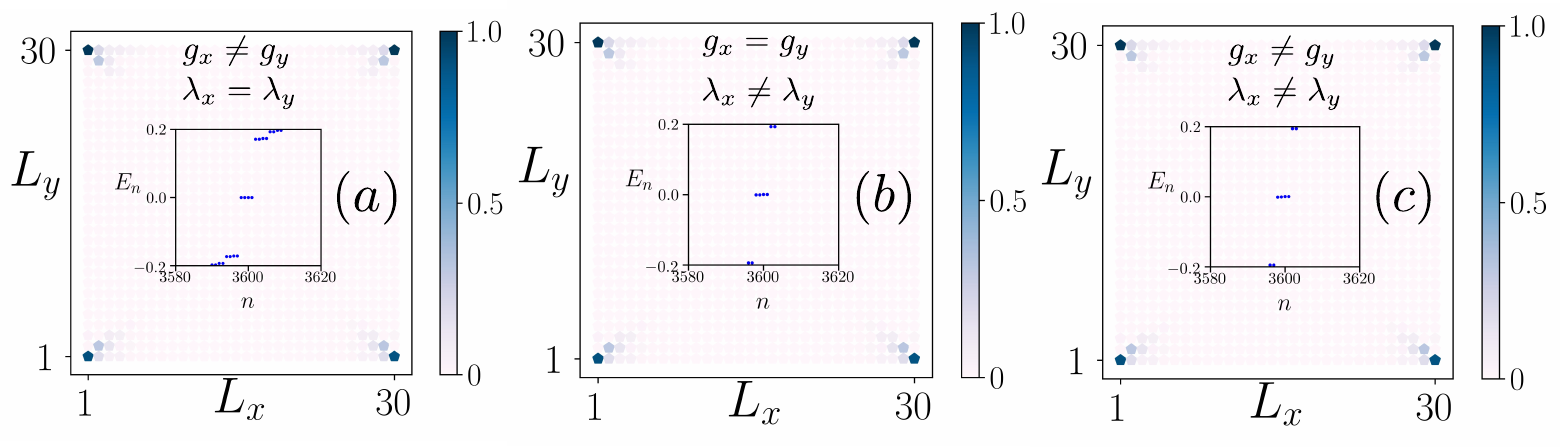}
	\end{center}
	\caption{\hspace*{-0.1cm}Panels (a), (b), and (c) correspond to the eigenvalue spectrum and the corresponding LDOS for three different scenarios. In panel (a), we consider $g_x = 0.2$, $g_y = 0.3$, 
	and $\lambda_x = \lambda_y = 0.5$. In panel (b), we take $g_x = g_y = 0.2$, and $\lambda_x = 0.5$, $\lambda_y = 0.8$. Finally, in panel (c), we choose $g_x = 0.2$, $g_y = 0.3$, and $\lambda_x = 
	0.5$, $\lambda_y = 0.8$. The rest of the model parameters remain the same as mentioned in the main text: $J_{ex} = 0.8$, $\Delta_0 = 0.4$, $\epsilon_0 = 1$.}
	\label{Fig1}
\end{figure}
\vspace{-1.0cm}
\section{Effect of Rashba spin-orbit coupling} \label{Sec:S5}
This section is devoted to the discussion of the effect of external Rashba SOC on SOTSC phase. In our theoretical setup, we propose a quantum spin hall insulating (QSHI) layer on top of a 2D non-collinear magnetic texture. In particular, the QSHI layer of our setup possesses an intrinsic SOC while the non-collinear magnetic texture (spin spiral) can generate another effective SOC. Therefore, SOTSC hosting MCMs are stabilized due to the interplay between these two types of SOC. Hence, we do not need any Rashba SOC to obtain the SOTSC. However, we here present a qualitative study 
of our problem in the presence of external Rashba SOC to examine the stability of the SOTSC phase. Therefore, the Hamiltonian in Eq.~(1) of the main text, in the presence of Rashba SOC, is modified to,

\begin{figure}[H]
	\begin{center}
		\includegraphics[width=0.4\textwidth]{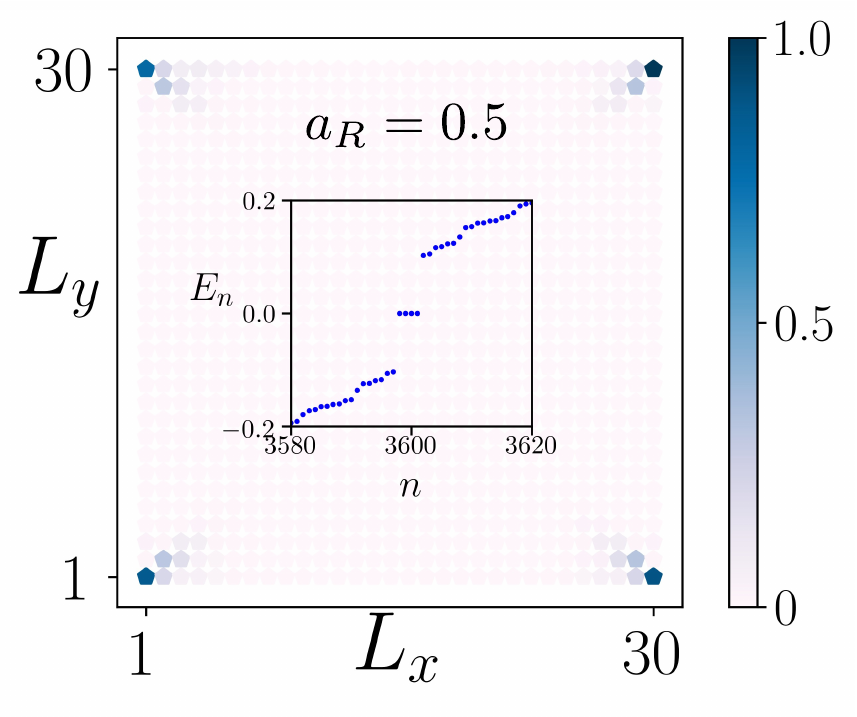}
	\end{center}
	\caption{\hspace*{-0.1cm} We illustrate the eigenvalue spectrum and the corresponding LDOS with the Rashba SOC strength $a_R=0.5$. The rest of the model parameter values remain the same 
	as mentioned in the main text: $J_{ex} = 0.8$, $\Delta_0 = 0.4$, $\epsilon_0 = 1$, $g_x = g_y=0.2$ and $\lambda_x = \lambda_y=0.5$.}
	\label{Fig2}
\end{figure}

\begin{align}
H &= \sum_{i,j} c_{i,j}^\dagger  \left\{ \epsilon_0 \Gamma_1 + \Delta_{0} \Gamma_2 + J_{\text{ex}} \left(\Gamma_3 \cos \phi_{ij} + \Gamma_4 \sin \phi_{ij} \right) \right\} c_{i,j} \nonumber \\
&\quad - c_{i,j}^\dagger\left\{ t \Gamma_1 + i \lambda_x \Gamma_5 - i\frac{a_R}{2} \Gamma_5^R \right\} c_{i+1,j} \nonumber \\
&\quad - c_{i,j}^\dagger\left\{ t \Gamma_1 + i \lambda_y \Gamma_6 + i\frac{a_R}{2} \Gamma_6^R \right\} c_{i,j+1}  + \text{h.c.} \ ,
\label{Eq37}
\end{align}
where, the lattice site indices $i$ and $j$ runs along $x$- and $y$-direction, respectively and the $\vect{\Gamma}$ matrices ($8 \times 8$) are given as  $\Gamma_1$=$\tau_z \sigma_z s_0$, $\Gamma_2$=$\tau_x\sigma_0 s_0$, $\Gamma_3$=$\tau_0\sigma_{0}s_x$, $\Gamma_4$=$\tau_0\sigma_{0}s_y$, $\Gamma_5$=$\tau_z\sigma_x s_z$, $\Gamma_6$=$\tau_z\sigma_y s_0$, $\Gamma_7$=$\tau_z\sigma_0 s_y$, and $\Gamma_8$=$\tau_z\sigma_0 s_x$. The three Pauli matrices $\vect{\sigma}, \vect{s}$ and $\vect{\tau}$ act on orbital ($a, b$), spin ($\uparrow, \downarrow$), and particle-hole degrees of freedom, respectively. The symbol $\lambda_R$ represents the strength of the Rashba SOC, and the rest of the symbols carry the same meaning as mentioned in the main text. In Fig. \ref{Fig2}, we depict the eigenvalue spectrum and the corresponding LDOS using Eq.~(\ref{Eq37}) choosing a moderate value of the Rashba SOC strength. We still obtain the SOTSC phase hosting MCMs at $E=0$ 
for $\lambda_R \neq 0$. Thus, the presence of Rashba SOC does not modify the presence of MCMs in our system. Therefore, we can conclude that in the presence of Rashba SOC, we can't expect any 
new physics. Instead, the Rashba SOC term $a_R$ merely can lead to a renormalization of certain topological regime.

\vspace{0.3cm}
\section{MCMs in Disc and Triangular Geometry} \label{Sec:S6}
Finally, in this section of the SM, we present our findings for both circular disc and triangular geometries based on our lattice model [Eq.~(1) in the main text]. In Fig.~\ref{Fig3}(a), we illustrate the LDOS 
and the corresponding eigenvalue spectrum (inset) for the circular disc geometry. In Figs.~\ref{Fig3}(b) and (c), we demonstrate the same employing triangular geometry for two different orientations. It is evident that zero-dimensional localized MCMs emerge in the LDOS spectrum at $E=0$ supported by the eigenvalue spectra. Notably, the critical distinction between the two geometries lies in the number 
of MCMs: the disc geometry supports four corner modes [see Fig.~\ref{Fig3}(a)] while the triangular setup only exhibits two corner modes [Figs.~\ref{Fig3}(b) and (c)]~\cite{Zhang2020,Trauzettel2020}. Interestingly, the position of these corner modes depends on the orientations of the triangle. This has also been shown in other models of SOTSC~\cite{Zhang2020,Trauzettel2020}. Therefore, our results affirm that higher-order topology remains consistent irrespective of the system's geometric configuration.

\vspace{0.5cm}
\begin{figure}[H]
	\begin{center}
		\includegraphics[width=1.0\textwidth]{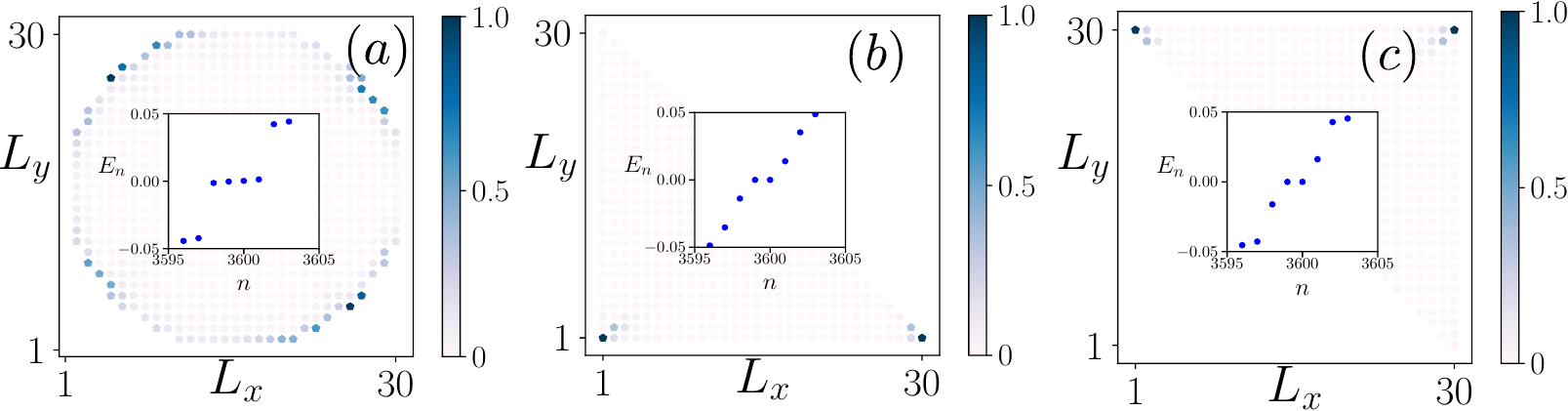}
	\end{center}
	\caption{In panel (a), we show the LDOS associated with $E=0$, for circular disc geometry. In panels (b) and (c), we showcase the same employing triangular geometry with two different orientations.  
	In the insets, we depict the eigenvalue spectra for the corresponding geometries highlighting the $E=0$ localized MCMs. We assume all the model parameters remain same as mentioned in the 
	main text \ie for the square geometry: $J_{ex} = 0.8$, $\Delta_0 = 0.4$, $\epsilon_0 = 1.0$, $\lambda_x = \lambda_y = 0.5$,  $g_x = g_y = 0.2$, and $t=1.0$.}
	\label{Fig3}
\end{figure} 
\end{onecolumngrid}
\end{document}